\documentclass[12pt]{iopart}
\usepackage{amsfonts}
\usepackage{bm}
\usepackage{amssymb}
\usepackage{epsfig}

\newcommand{\kM}{{\bm\mu}}

\begin{document}
\title{Localised control for non-resonant Hamiltonian systems}

\author{M Vittot, C Chandre, G Ciraolo, R Lima}
\address{CPT-CNRS, Luminy Case 907, F-13288 Marseille Cedex 9,
France}
\eads{\mailto{chandre@cpt.univ-mrs.fr}, \mailto{ciraolo@cpt.univ-mrs.fr}, \mailto{lima@cpt.univ-mrs.fr}, \mailto{vittot@cpt.univ-mrs.fr}}

\begin{abstract}
We present a method of {\em localised} control of chaos in Hamiltonian systems. The aim is to modify the perturbation {\em locally} by a small control term which makes the controlled Hamiltonian more regular. We provide an explicit expression for the control term which is able to recreate invariant (KAM) tori without modifying other parts of phase space.  
We apply this method of localised control to a forced pendulum model, to the delta-kicked rotor (standard map) and to a non-twist Hamiltonian.
\end{abstract}


\maketitle

\section{Introduction}

Controlling chaotic transport is a key challenge in many branches of physics like for instance, in particle accelerators, free electron lasers or in magnetically confined fusion plasmas.
One way to control transport would be that of reducing or
suppressing chaos. There exist numerous attempts to control chaos (see Refs.~\cite{review1,review2} for a rather extended list of references). Most of the methods for controlling chaotic systems is done by tilting targeted trajectories. However, for many body experiments like the magnetic confinement of a plasma or the control of turbulent flows, such methods are hopeless due to the high number of trajectories to deal with simultaneously. For these systems, it is desirable to control transport properties without significantly
altering the original system under investigation nor its overall chaotic structure. Here we focus on another strategy which is based on building barriers by adding a small apt perturbation which is localised in phase space, hence confining all the trajectories.

The main motivations for a {\em localised control} are the following ones~: Very often the control of a physical system can only be performed in some specific regions of phase space. This is in particular the case in thermonuclear fusion devices where the electric potential can only be modified near the border of the plasma. For some purposes it is sometimes desirable to stabilize only a given region of phase space without modifying the major part of phase space in order to preserve some specific features of the system. This method can be used to bound the motion of particles without changing the perturbation inside (and outside) the barrier. Also, using a localised control means that one needs to inject much fewer energy than a global control in order to create isolated barriers of transport.

In this article, we consider the class of Hamiltonian systems that can be written
in the form $H=H_0+\varepsilon V$ i.e.\ an integrable 
Hamiltonian $H_0$ (with action-angle variables)
plus a small perturbation $\varepsilon V$. The idea is to {\em slightly} and {\em locally} modify the perturbation and create regular structures (like invariant tori)~: The aim is to devise a control term $f$ such that
the dynamics of the controlled Hamiltonian $H_c=H_0+\varepsilon V+f$
has more regular trajectories or less diffusion
than the uncontrolled one. For practical purposes, the control term should be small
with respect to the perturbation $\varepsilon V$, and localised in phase space
(i.e.\ the subset of phase space where $f$ is non-zero is finite
and small).

In Refs.~\cite{guido1,guido2,michel}, an explicit method of control was provided in order to construct a control term $f$ of order $\varepsilon^2$ such that the controlled Hamiltonian $H_0+\varepsilon V+f$ is integrable. The main drawback of this approach is that the control term has to be applied on all the phase space.
Here we provide a method to construct control terms $f$ of order $\varepsilon^2$ with a finite support in phase space, such that the controlled Hamiltonian $H_c=H_0+\epsilon V+ f$ has isolated invariant tori. For Hamiltonian systems with two degrees of freedom, these invariant tori act as barriers in phase space. For higher dimensional systems KAM tori act as effective barriers of diffusion.

The main result of the paper is stated as follows~: For a Hamiltonian system written in action-angle variables with $L$ degrees of freedom, the perturbed Hamiltonian is $H({\bf A},{\bm\theta})={\bm \omega}\cdot {\bf A}+ V({\bf A},{\bm\theta})$ where $({\bf A},{\bm\theta})\in {\mathbb R}^L\times {\mathbb T}^L$ and $\bm\omega$ is a non-resonant vector of ${\mathbb R}^L$. 
We consider a region near ${\bf A}={\bf 0}$ and the perturbation $V$ has constant and linear parts in actions of order $\varepsilon$, i.e.\ $V({\bf A},{\bm\theta})=\varepsilon v({\bm\theta})+\varepsilon {\bf w}({\bm\theta})\cdot {\bf A}+Q({\bf A},{\bm\theta})$ where $Q$ is of order $O(\Vert {\bf A}\Vert ^2)$. We notice that for $\varepsilon=0$, the Hamiltonian $H$ has an invariant torus with frequency vector ${\bm\omega}$ at ${\bf A}={\bf 0}$ for any $Q$ not necessarily small.
The controlled Hamiltonian we construct is
\begin{equation}
\label{eqn:gene}
H_c({\bf A},{\bm\theta})={\bm \omega}\cdot {\bf A}+ V({\bf A},{\bm\theta})+  f({\bm \theta})\Omega(\Vert {\bf A}\Vert ),
\end{equation}
where $\Omega$ is a smooth characteristic function of a region around a targeted invariant torus (the size of its support is of order $\varepsilon$). It is sufficient to have $\Omega(\Vert {\bf A}\Vert)=1$ for $\Vert {\bf A}\Vert \leq \varepsilon$. For instance, $\Omega=1$ would be a possible and simpler candidate, however representing a long-range control. We notice that the control term $f$ we construct only depends on the angle variables and is given by
\begin{equation}
\label{eqn:exf}
f({\bm\theta})=V({\bf 0},{\bm\theta})-V\left( -\Gamma \partial_{\bm\theta} V({\bf 0},{\bm\theta}),{\bm\theta}\right),
\end{equation}
where $\Gamma$ is a linear operator defined below as a pseudo-inverse of ${\bm\omega}\cdot \partial_{\bm\theta}$. Note that $f$ is of order $\varepsilon^2$. For a sufficiently small perturbation, Hamiltonian~(\ref{eqn:gene}) has an invariant torus with frequency vector close to ${\bm\omega}$. After proving this result, we check numerically that the controlled Hamiltonian is more regular than the uncontrolled one, i.e.\ the invariant tori of the controlled Hamiltonian persist to higher values of the amplitude of the perturbation than in the uncontrolled case.

In Sec.~\ref{sec:1}, we explain the theory of the localised control of Hamiltonian systems and in particular we prove Eqs.~(\ref{eqn:gene})-(\ref{eqn:exf}). In Sec.~\ref{sec:2}, we give some applications of the localised control on the following models: a forced pendulum Hamiltonian, the delta-kicked rotor (standard map) and a non-twist Hamiltonian model. For these systems, we show numerically that the localised control is able to create isolated invariant tori beyond the values of the parameters for which there are no invariant tori in the absence of control.

\section{Localised control for Hamiltonian systems}
\label{sec:1}

\subsection{Global control~: toward a localised control theory}
We first recall the global control theory as explained in Refs.~\cite{michel,guido2} in order to define the main operators that will be used for the localised control. \\
Let us fix a Hamiltonian $H_0$.
We define the linear operator $\{H_0\}$ by
$$
\{H_0\}H=\{H_0,H\},
$$
where $\{\cdot ,\cdot\}$ is the Poisson bracket.
 The operator
\(\{{H_0}\} \) is not invertible, e.g., $\{H_0\}H_0=0$. We consider a pseudo-inverse of \( \{{H_0}\} \), denoted by $\Gamma$, satisfying
\begin{equation}
\{{H_0}\}^{2}\ \Gamma = \{{H_0}\}.
\label{gamma}
\end{equation}
If the operator $\Gamma$ exists, it is not unique in general. We define the {\em resonant} operator $\mathcal R$ as
\begin{equation}
{\mathcal R} = 1-\{H_0\}\Gamma,
\end{equation} 
We notice that Eq.~(\ref{gamma}) becomes $\{{H_0}\} \mathcal R = 0$.
A consequence
is that any element ${\mathcal R} V$ is constant under the flow of $H_0$. 

{\em Notation~:} In what follows, we will use the notation ${\bf a} {\bf b}$ for an operation between ${\bf a}$ and ${\bf b}$ which can be vectors or covectors. For instance, if ${\bf a}$ is a covector and ${\bf b}$ a vector, ${\bf a} {\bf b}$ is the usual scalar product ${\bf a}\cdot {\bf b}$. If ${\bf a}$ is a vector and ${\bf b}$ a covector, ${\bf a} {\bf b}$ is the matrix whose elements are $[{\bf a} {\bf b}]_i^j= a_i b^j$. For a vector ${\bf a}$ and a matrix $M$, $M {\bf a}$ is a vector. In the same way, if ${\bf a}$ is a covector, ${\bf a} M$ is a covector. Also we denote ${\bf a}'$ the matrix $\partial_{\bm\theta} {\bf a}$ with elements $[\partial_{\bm\theta} {\bf a}]_i^j=\partial a^j/\partial \theta^i$. For clarity we also denote $\partial$ the operator $\partial_{\bm\theta}$.\\

Let us now assume that $H_0$ is integrable with action-angle variables 
$({\bf A},\bm{\theta})\in {\mathbb R}^L\times \mathbb{T}^L $ where ${\mathbb T}^L$ is the $L$-dimensional torus. Here, ${\bf A}$ is a $L$-dimensional vector and ${\bm \theta}$ is a $L$-dimensional covector. The Poisson bracket between two functions $H$ and $V$ is given in the usual form
$$
\{H,V\}=\frac{\partial H}{\partial {\bf A}} \frac{\partial V}{\partial {\bm\theta}}-
\frac{\partial V}{\partial {\bf A}} \frac{\partial H}{\partial {\bm\theta}}.
$$
We assume that $H_0$ is linear in the actions variables, so that $H_0={\bm \omega} {\bf A}$, where the frequency vector ${\bm \omega}$ is any co-vector of ${\mathbb R}^L$. In this paper, we assume that ${\bm \omega}$ is non-resonant, i.e.\ there is no vector ${\bf k}\in {\mathbb Z}^L\setminus \{ {\bf 0}\}$ such that ${\bm\omega}{\bf k}=0$.
The operator $\{H_0\}$ acts on $V$ given by
$$
V=\sum_{{\bf k}\in {\mathbb Z}^L}V_{\bf k}({\bf A}){\mathrm e}^{i{\bm\theta} {\bf k}},
$$
as
$$
(\{H_0\}V)({\bf A},\bm{\theta})=\sum_{\bf k\in {\mathbb Z}^L}i{\bm \omega} {\bf k}~V_{\bf k}({\bf A}){\mathrm e}^{i{\bm\theta}{\bf k}}.
$$
A possible choice of $\Gamma$ is
$$
(\Gamma V)({\bf A},\bm{\theta})=
\sum_{{\bf k}\in{\mathbb Z^L}\atop{{\bm \omega}{\bf k}\neq0}}
\frac{V_{\bf k}({\bf A})}
{i{\bm \omega} {\bf k}}~~{\mathrm e}^{i{\bm\theta}{\bf k}}.
$$
We notice that this choice of $\Gamma$ commutes with $\{H_0\}$.
\\ \indent The operator ${\mathcal R}$ is the projector on the resonant 
part of the perturbation:
\begin{equation}
{\mathcal R}V=\sum_{{\bm \omega}{\bf k}=0}
V_{\bf k}({\bf A}){\mathrm e}^{i{\bm\theta}{\bf k}}= V_{\bf 0} ({\bf A}),\label{eqn:RV}
\end{equation}
since ${\bm \omega}$ is non-resonant.
We also define the projector on the non-resonant part of the perturbation 
\begin{equation}
{\mathcal N}V=\sum_{{\bm \omega}{\bf k}\not= 0}
V_{\bf k}({\bf A}){\mathrm e}^{i{\bm\theta}{\bf k}}=V-V_{\bf 0}.\label{eqn:NV}
\end{equation}
The global control follows directly from the definition of these operators $\Gamma$, ${\mathcal R}$ and ${\mathcal N}$~: We construct a global control term for the perturbed Hamiltonian $H_0+V$, i.e.\ we construct $f$ such that the controlled Hamiltonian $H_c=H_0+V+f$
is canonically conjugate to $H_0+\mathcal R V$. This conjugation is given by the following equation
\begin{equation}
{\mathrm e}^{\{\Gamma V\}}(H_0+V+f)=H_0+{\mathcal R} V,
\label{prop1}
\end{equation}
where
\begin{equation}
\label{eqn:fctf}
f(V)=\sum_{n=1}^{\infty}\frac{(-1)^n}{(n+1)!}\{\Gamma V\}^n
(n{\mathcal R}+1)V.
\label{eqn:ctf}
\end{equation}
We notice that
if $V$ is of order $\varepsilon$, the control term $f$ is of order $\varepsilon^2$. In general, the control term depends on all the variables ${\bf A}$ and ${\bm\theta}$, and acts {\em globally} on all phase space. 

Since ${\bm \omega}$ is non-resonant, ${\mathcal R}V=V_{\bf 0}({\bf A})$ only depends on the actions and thus $H_0+{\mathcal R}V$ is integrable. The derivation of Eqs.~(\ref{prop1})-(\ref{eqn:ctf}) is given in Refs.~\cite{michel,guido2}. 

Starting from this global control, we derive a localised control such that the control term only acts in a given region of phase space around a selected invariant torus. We consider a nearly integrable Hamiltonian system~:
\begin{equation}
\label{eqn:H0V}
H({\bf A},{\bm \theta})=H_0({\bf A})+ V({\bf A},{\bm \theta}).
\end{equation}
We assume that $H_0$ has the invariant torus with a non-resonant frequency vector ${\bm \omega}$ at ${\bf A}={\bf A}_0$.
For $V$ sufficiently small, the KAM theorem ensures that this invariant torus is preserved under suitable hypothesis. We expand Hamiltonian~(\ref{eqn:H0V}) around ${\bf A}={\bf A}_0$ and we translate the actions such that the invariant torus with frequency ${\bm\omega}$ is located at ${\bf A}={\bf 0}$ for $H_0$, and around ${\bf A}={\bf 0}$ for the perturbed Hamiltonian. Hamiltonian~(\ref{eqn:H0V}) becomes (up to a constant)
\begin{equation}
\label{eqn:Hcf}
H({\bf A},{\bm \theta})={\bm\omega}  {\bf A}+\varepsilon v({\bm \theta}) +\varepsilon {\bf w}({\bm \theta}) {\bf A} + Q({\bf A},{\bm \theta}),
\end{equation}
where $Q$ is of order $O(\Vert {\bf A}\Vert ^2)$, i.e., $Q({\bf 0},{\bm \theta})=0$ and $\partial_{\bf A}Q({\bf 0},{\bm \theta})={\bf 0}$.
Without any restriction, we assume that Hamiltonian~(\ref{eqn:Hcf}) is such that ${\mathcal R}v=0$ and ${\mathcal R} {\bf w}={\bf 0}$~: The mean value of $v$ is absorbed into the total energy and the mean value of ${\bf w}$ into the frequency vector ${\bm \omega}$. 

For ${\bf A}$ sufficiently small, the perturbation
$$
V({\bf A},{\bm\theta})=\varepsilon v({\bm \theta}) +\varepsilon {\bf w}({\bm \theta}) {\bf A} + Q({\bf A},{\bm \theta}),
$$
is small. We apply Eq.~(\ref{eqn:fctf}) in order to get the control term $f$. However, for larger ${\bf A}$, the control term is no longer small. Therefore we localise it in a region close to ${\bf A}={\bf 0}$, i.e.\ we consider the following controlled Hamiltonian~:
$$
H_c({\bf A},{\bm \theta})={\bm\omega}  {\bf A}+V({\bf A},{\bm\theta})+f({\bf A},{\bm\theta})\Omega(\Vert {\bf A}\Vert ),
$$
where $\Omega$ is a smooth characteristic function such that $\Omega(x)=0$ if $x\geq 2\varepsilon$, and $\Omega(x)=1$ if $x\leq \varepsilon$.
The main drawback of this approach is that the control term is a priori of order $\varepsilon$ even if it is small since it is localised in a region near ${\bf A}={\bf 0}$. In the next section we develop another approach where the control term $f$ is of order $\varepsilon^2$ and does no longer depend on ${\bf A}$.

\subsection{Localised control theory}
\label{sec:2b}
 
As in the previous section, we consider the family of Hamiltonians~(\ref{eqn:Hcf}). For $\varepsilon =0$, Hamiltonian~(\ref{eqn:Hcf}) has an invariant torus with frequency vector ${\bm \omega}$ located at ${\bf A}={\bf 0}$.
 The problem of control we address is to slightly modify Hamiltonian~(\ref{eqn:Hcf}) near ${\bf A}={\bf 0}$ in the following way~:
\begin{equation}
	H_c({\bf A},{\bm \theta})={\bm\omega}  {\bf A}+\varepsilon v({\bm \theta}) +\varepsilon {\bf w}({\bm \theta}) {\bf A}+ Q({\bf A},{\bm \theta}) +\varepsilon^2  f({\bm \theta})\Omega(\Vert {\bf A}\Vert),
\end{equation}
such that the invariant torus with frequency ${\bm \omega}$ exists for the controlled Hamiltonian $H_c$ for higher values of the parameter $\varepsilon$ than in the uncontrolled case. Here $\Omega$ denotes a smooth step function, meaning that the control only applies in a small part of phase space (of size $\varepsilon$)~: For instance, $\Omega$ is a sufficiently smooth function such that $\Omega(x)=1$ if $x\leq\varepsilon$ and $\Omega(x)=0$ if $x\geq 2\varepsilon$. Moreover, we notice that the control term $f$ we apply is only a function of the angles in the region around the invariant torus.

The main proposition of the {\em localised} control of Hamiltonian systems is the following one~:
 
{\em Proposition~1:} If $v$ and ${\bf w}$ are sufficiently small and if $v$, ${\bf w}$ and $Q$ are smooth, there exists a control term $f$ such that the controlled Hamiltonian
\begin{equation}
\label{eqn:HC}
	H_c({\bf A},{\bm \theta})={\bm\omega}  {\bf A}+\varepsilon v({\bm \theta}) +\varepsilon {\bf w}({\bm \theta}) {\bf A} + Q({\bf A},{\bm \theta})+\varepsilon^2  f({\bm \theta})\Omega(\Vert {\bf A}\Vert),
\end{equation}
is canonically conjugate to
\begin{equation}
\label{eqn:htildc}
	\tilde{H}_c({\bf A},{\bm \theta})=\tilde{\bm\omega}  {\bf A}+ \tilde{Q}({\bf A},{\bm \theta}),
\end{equation}
where $\tilde{\bm\omega}={\bm\omega}+\varepsilon {\bf a}$ with a constant covector ${\bf a}$ and $\tilde{Q}$ is of order $O(\Vert {\bf A}\Vert ^2)$, i.e.\ $\tilde{Q}({\bf 0},{\bm \theta})=0$ and $\partial_{\bf A} \tilde{Q}({\bf 0},{\bm \theta})={\bf 0}$. 
The control term $f$ is given by
\begin{equation}
\label{eqn:CTh}
	f({\bm \theta})={\mathcal N} \left[ {\bf w} \Gamma \partial v-\varepsilon^{-2}Q(-\varepsilon \Gamma \partial v,{\bm \theta})\right],
\end{equation}
which can also be written as 
$$
f({\bm \theta})=\varepsilon^{-2} {\mathcal N} \left[ V({\bf 0},{\bm\theta})-V\left( -\Gamma \partial V({\bf 0},{\bm\theta}),{\bm\theta}\right) \right].
$$
The important feature of the control term $f$ is that it does only depend on the angle variables.
Since the Hamiltonian $\tilde{H_c}$ has an invariant torus with frequency $\tilde{\bm\omega}$ at ${\bf A}={\bf 0}$, the controlled Hamiltonian $H_c$ has also this invariant torus in the region where ${\bf A}$ is close to ${\bf 0}$. 

{\em Proof:} We consider the following transformations $T_{{\bf m},{\bf b}}$ acting on functions $V({\bf A},{\bm \theta})$ like
$$
(T_{{\bf m},{\bf b}}V)({\bf A},{\bm \theta})={\mathrm e}^{-{\bf m} \partial} \left[ V({\mathrm e}^{\partial \hat{\bf m}}{\bf A} +{\bf b},{\bm \theta})\right],
$$ 
where the covector ${\bf m}$ and the vector ${\bf b}$ are functions of ${\bm \theta}$ from ${\mathbb T}^L$ into ${\mathbb R}^L$. The operator $\hat{{\bf m}}$ is the linear operator from ${\mathbb R}^L$ to ${\mathbb R}$ acting on a vector ${\bf u}\in {\mathbb R}^L$ as $\hat{{\bf m}}{\bf u}= \bf m {\bf u}$. Here $\partial \hat{\bf m}$ is the linear operator from ${\mathbb R}^L$ to ${\mathbb R}^L$ which is the product of the two linear operators acting on a vector ${\bf u}\in {\mathbb R}^L$ as
\begin{equation}
\label{eqn:dmu}
\partial \hat{\bf m} {\bf u}={\bf m}'{\bf u} + {\bf m} {\bf u}'.
\end{equation}

In \ref{sec:appa}, we check that the transformations $T_{{\bf m},{\bf b}}$ are canonical if ${\bf b}$ derives from a scalar function. 
We perform a transformation $T_{{\bf m},{\bf b}}$ on the controlled Hamiltonian $H_c$ given by Eq.~(\ref{eqn:HC}) and we determine the functions ${\bf m}$ and ${\bf b}$ in the following ways~: 

\noindent $(i)$ The function ${\bf b}$ is determined such that the order $\varepsilon$ of the constant term in actions vanishes.\\
$(ii)$ The function ${\bf m}$ is determined such that the linear term in actions (which is of order $\varepsilon$) vanishes.\\
$(iii)$ The control term $f$ is determined such that the constant term in actions [which is now of order $\varepsilon^2$ after $(i)$] vanishes.

We perform a transformation $ T_{\varepsilon{\bf m},\varepsilon{\bf b}} $ which is $\varepsilon$-close to the identity~: 
The expression of $\tilde{H}_c=T_{\varepsilon{\bf m},\varepsilon{\bf b}} H_c$ is
\begin{eqnarray}
	\tilde{H}_c=&&{\mathrm e}^{-\varepsilon {\bf m} \partial } \Bigl[ {\bm\omega} {\bf A}+\varepsilon {\bm \omega} {\bf b}+\varepsilon v({\bm \theta})\nonumber \\
	 && +\varepsilon {\bm\omega}  \kM'\, {\bf A}+\varepsilon {\bf w} {\bf A} + Q((1+\varepsilon  \kM '){\bf A}+\varepsilon {\bf b},{\bm \theta})\nonumber \\ && +\varepsilon^2 {\bf w}\kM' {\bf A}+\varepsilon^2 {\bf w} {\bf b} +\varepsilon^2 f({\bm \theta})\Omega(\Vert (1+\varepsilon \kM'){\bf A}+\varepsilon {\bf b} \Vert)  \Bigr],
\end{eqnarray}
where the covector $\kM$ is defined by
\begin{equation}
\label{eqn:mu}
\kM =\frac{{\mathrm e}^{\varepsilon {\bf m}\partial}-1}{ \varepsilon {\bf m}\partial} {\bf m}=\sum_{n=0}^\infty \frac{\left( \varepsilon {\bf m}\partial\right)^n}{(n+1)!}{\bf m}.
\end{equation}
The function $\kM$ is $\varepsilon$-close to ${\bf m}$~:
$$
\kM={\bf m}+\varepsilon {\bf m} {\bf m}'+\cdots,
$$
and satisfies
$$
{\mathrm e}^{\varepsilon \partial \hat{\bf m}}{\bf A}={\bf A}+\varepsilon  \kM' {\bf A},
$$
so that 
\begin{equation}
\label{eqn:wf}
1+\varepsilon  \kM' ={\mathrm e}^{\varepsilon \partial \hat{\bf m}}\cdot 1,
\end{equation}
where ${\mathrm e}^{\varepsilon \partial \hat{\bf m}}\cdot 1 $ is a matrix, function of ${\bf \theta}$, which results from the action of the operator ${\mathrm e}^{\varepsilon \partial \hat{\bf m}}$ on the constant function 1.
First we notice that the scalar function
$$
q({\bf A},{\bm \theta})=1-\Omega\left(\Vert (1+\varepsilon \kM'){\bf A}+\varepsilon {\bf b}\Vert\right),
$$
is of order $O(\Vert {\bf A}\Vert^2)$. This can be seen from the equations
\begin{eqnarray*}
	&& q({\bf 0},{\bm \theta})=1-\Omega\left(\varepsilon \Vert{\bf b}\Vert\right),\\
	&& \partial_{\bf A} q ({\bf 0},{\bm \theta})=-\Omega'\left(\varepsilon\Vert{\bf b}\Vert\right)\frac{\bar{\bf b}}{\Vert{\bf b}\Vert} (1+\varepsilon \kM'), 
\end{eqnarray*}
where $\bar{\bf b}$ is the transposed covector of ${\bf b}$.
The function ${\bf b}({\bm \theta})$ will be chosen such that $\Vert {\bf b}\Vert \leq 1$ for all ${\bm \theta}$. Therefore we have 
$q({\bf 0},{\bm \theta})=0$ and $\partial_{\bf A} q ({\bf 0},{\bm \theta})={\bf 0}$, according to the hypothesis on the smooth step function $\Omega$. \\
Next, we expand the function $Q((1+\varepsilon \kM'){\bf A}+\varepsilon {\bf b},{\bm \theta})$ around ${\bf A}={\bf 0}$~:
$$
Q((1+\varepsilon  \kM'){\bf A}+\varepsilon {\bf b},{\bm \theta})=Q(\varepsilon {\bf b},{\bm \theta})+\partial_{\bf A} Q(\varepsilon {\bf b},{\bm \theta}) (1+\varepsilon  \kM') {\bf A}+ Q_2({\bf A},{\bm\theta}),
$$
where $Q_2$ is of order $O(\Vert {\bf A} \Vert ^2)$, i.e.\ $Q_2({\bf 0},{\bm \theta})=0$ and $\partial_{\bf A} Q_2({\bf 0},{\bm \theta})={\bf 0}$.
We notice that $Q(\varepsilon {\bf b},{\bm \theta})$ is of order $\varepsilon^2$ and the covector $\partial_{\bf A}Q(\varepsilon {\bf b},{\bm \theta})$ is of order $\varepsilon$ since $Q$ is of order $O(\Vert {\bf A}\Vert^2)$.
The Hamiltonian $\tilde{H_c}$ becomes
\begin{eqnarray}
	\tilde{H}_c=&&{\mathrm e}^{-\varepsilon {\bf m} \partial} \Bigl[ {\bm\omega}  {\bf A}+\varepsilon {\bm \omega} {\bf b}+\varepsilon v({\bm \theta}) \nonumber \\
	 && +\varepsilon {\bm\omega} \kM' \, {\bf A}+\varepsilon {\bf w} {\bf A} +\varepsilon^2 {\bf w}\kM' \, {\bf A} + \partial_{\bf A}Q(\varepsilon {\bf b},{\bm \theta}) (1+\varepsilon  \kM'){\bf A} \nonumber \\ && +Q(\varepsilon {\bf b},{\bm \theta}) +\varepsilon^2 {\bf w} {\bf b} +\varepsilon^2 f({\bm \theta})
	 +Q_2({\bf A},{\bm\theta})-f({\bm \theta})q({\bf A},{\bm\theta})\Bigr].
\end{eqnarray}

The canonical transformation is determined by two equations~:
\begin{eqnarray}
	&& {\bm \omega} {\bf b}+v=0, \label{eqn:detb}\\
	&& {\mathcal N} \left[ ({\bm \omega}+\varepsilon {\bf w} +\partial_{\bf A}Q(\varepsilon {\bf b},{\bm\theta})) \kM' +{\bf w}+\varepsilon^{-1} \partial_{\bf A} Q(\varepsilon {\bf b},{\bm\theta})\right]={\bf 0}. \label{eqn:detm}
\end{eqnarray}
The control term is chosen such that 
$$
f({\bm \theta})=-{\mathcal N} \left[ {\bf w} {\bf b} +\varepsilon^{-2} Q(\varepsilon {\bf b},{\bm \theta})\right].
$$
Equations~(\ref{eqn:detb}) is solved in Fourier space. We expand the function ${\bf b}$~:
$$
{\bf b}({\bm \theta})=\sum_{{\bf k}\in {\mathbb Z}^L} {\bf b}_{\bf k} {\mathrm e}^{i{\bm \theta}{\bf k} },
$$
and the coefficients ${\bf b}_{\bf k}$ are given according to Eq.~(\ref{eqn:detb})~:
$$
{\bf b}_{\bf k}=\frac{\bf k}{{\bm \omega}{\bf k}} v_{\bf k},
$$
when ${\bm \omega} {\bf k}\not= 0$, and ${\bf b}_{\bf k}={\bf 0}$ when ${\bm \omega} {\bf k}= 0$. Thus the vectorial function ${\bf b}$ is chosen to be
$$
{\bf b}=-\Gamma \partial v.
$$
We recall that we require that $\Vert {\bf b} \Vert=\sup_{\bm\theta} \vert {\bf b}\vert \leq 1$ which is ensured if $v$ is sufficiently small and smooth and if ${\bm\omega}$ satisfies a Diophantine condition (see the usual KAM proofs like for instance in Ref.~\cite{kam}).
In particular, we notice that this choice of ${\bf b}$ satisfies ${\mathcal R} {\bf b}={\bf 0}$.
Equation~(\ref{eqn:detm}) is solved by choosing $\kM=\Gamma \tilde{\kM}$, where $\tilde{\kM}$ satisfies 
\begin{equation}
\label{eqn:muti}
\tilde{\kM}+(\varepsilon {\bf w} +\partial_{\bf A}  Q(-\varepsilon \Gamma \partial v,{\bm\theta})) \Gamma \tilde{\kM}' +{\bf w}+\varepsilon^{-1} \partial_{\bf A}  Q(-\varepsilon \Gamma \partial v,{\bm\theta})={\bf 0}.
\end{equation}
It is straightforward to check that Eq.~(\ref{eqn:detm}) is satisfied if $\tilde{\bm\mu}$ is a solution of Eq.~(\ref{eqn:muti}). If the operator $1+(\varepsilon {\bf w}+\partial_{\bf A} Q(-\varepsilon \Gamma \partial v,{\bm \theta})) \Gamma \partial $, which is $\varepsilon$-close to the identity and then invertible, Eq.~(\ref{eqn:detm}) has a solution
$$
\kM=-\Gamma \left[ 1+(\varepsilon {\bf w}+\partial_{\bf A} Q(-\varepsilon \Gamma \partial v,{\bm \theta})) \Gamma \partial \right]^{-1} ({\bf w}+\varepsilon^{-1}\partial_{\bf A} Q(-\varepsilon \Gamma \partial v,{\bm \theta})).
$$
The hypothesis on invertibility is fulfilled if ${\bf w}$ as well as $v$ are small enough and smooth and if ${\bm\omega}$ satisfies a Diophantine condition (see again Ref.~\cite{kam}).
The covector $\kM$ which is $\varepsilon$-close to ${\bf m}$ has the following expansion
$$
\kM=-\Gamma {\bf w}-\varepsilon^{-1}\Gamma \partial_{\bf A} Q(-\varepsilon\Gamma\partial  v,{\bm\theta})  +O(\varepsilon).
$$
We notice that ${\mathcal R}\kM={\bf 0}$ since $\Gamma$ and ${\mathcal R}$ commute.
The resulting Hamiltonian is then given by Eq.~(\ref{eqn:htildc}) where $\tilde{\bm\omega}={\bm\omega}+\varepsilon {\mathcal R} \tilde{\kM}$ and $\tilde{Q}={\mathrm e}^{-\varepsilon {\bf m}\partial}(Q_2-qf)$ which is of order $O(\Vert {\bf A}\Vert ^2)$. Note that
we have dropped some additive constants of order $\varepsilon^2$, ${\mathcal R} Q(-\varepsilon \Gamma \partial v,{\bm \theta})$ and ${\mathcal R} ({\bf w} \Gamma \partial v)$ since we recall that ${\mathcal R}$ is the mean value with respect to the angles.
The equations of motion for $\tilde{H}_c$ are
\begin{eqnarray*}
 && \dot{\bm \theta}=\tilde{\bm\omega}+\partial_{\bf A} \tilde{Q}({\bf A},{\bm\theta}),\\
 && \dot{\bf A}=-\partial_{\bm\theta}\tilde{Q}({\bf A},{\bm\theta}).
\end{eqnarray*}
Since $\tilde{Q}({\bf 0},{\bm \theta})=0$ and $\partial_{\bf A} \tilde{Q}({\bf 0},{\bm \theta})={\bf 0}$, we see that ${\bf A}={\bf 0}$ is invariant, and that the evolution of the angles is linear in time with frequency vector $\tilde{\bm\omega}$.
Therefore, the Hamiltonian $\tilde{H}_c$ has an invariant torus located at ${\bf A}={\bf 0}$ with frequency vector $\tilde{\bm\omega}$. 
More precisely, the flow of the controlled Hamiltonian $H_c$ on ${\bf A}={\bf 0}$ is
$$
{\mathrm e}^{t\{H_c\}} T_{\varepsilon{\bf m},\varepsilon{\bf b}}^{-1} \left(\begin{array}{c} {\bf 0} \\{\bm\theta}\end{array}\right) =   T_{\varepsilon{\bf m},\varepsilon{\bf b}}^{-1} \left(\begin{array}{c} {\bf 0} \\ {\bm\theta}+\tilde{\bm\omega} t\end{array} \right),
$$
where 
$$
T_{\varepsilon{\bf m},\varepsilon{\bf b}}^{-1} \left( \begin{array}{c} {\bf A} \\ {\bm\theta} \end{array}\right) = \left( \begin{array}{c}  {\mathrm e}^{-\varepsilon \partial  \hat{\bf m}} ({\bf A}-{\bf b}) \\ {\bm \theta}+\varepsilon {\bf m}\end{array} \right).
$$
The equation of the torus ${\bf A}={\bf 0}$ is thus 
$$
T_{\varepsilon{\bf m},\varepsilon{\bf b}}^{-1} \left( \begin{array}{c} {\bf 0} \\ {\bm\theta} \end{array}\right) = \left( \begin{array}{c}  -{\mathrm e}^{-\varepsilon \partial  \hat{\bf m}}{\bf b} \\ {\bm \theta}+\varepsilon {\bf m}\end{array} \right).
$$ 
\hfill $\Box$ \\

{\em Remark 1~:}
We notice that if $v=0$, the control term $f$ given by Eq.~(\ref{eqn:CTh}) is zero. In this case, the original Hamiltonian already has the invariant torus at ${\bf A}={\bf 0}$.\\

{\em Remark 2~: Addition property of the control term--} In the case where more than one invariant torus needs to be created, we can add the control terms localised in non-overlapping regions of phase space. This is a straightforward extension to the previous case. The controlled Hamiltonian becomes
$$
H({\bf A},{\bm \theta})=H_0({\bf A})+\varepsilon V({\bf A},{\bm \theta})+\varepsilon^2 \sum_{i=1}^M f_i({\bm\theta})\Omega\left(\Vert {\bf A}-{\bf A}_i\Vert\right) ,
$$
where $f_i$ is defined for each region of phase space by Eq.~(\ref{eqn:CTh}). We notice that in each region of phase space, the operators $\Gamma$, ${\mathcal R}$ and ${\mathcal N}$ are different since they are defined from the frequency vector of a given invariant torus.

\section{Applications}
\label{sec:2}
\subsection{Forced pendulum Hamiltonian}

\begin{figure}
\unitlength 1cm
\begin{picture}(7.5,6)(0,0)
\put(0,0){\epsfig{file=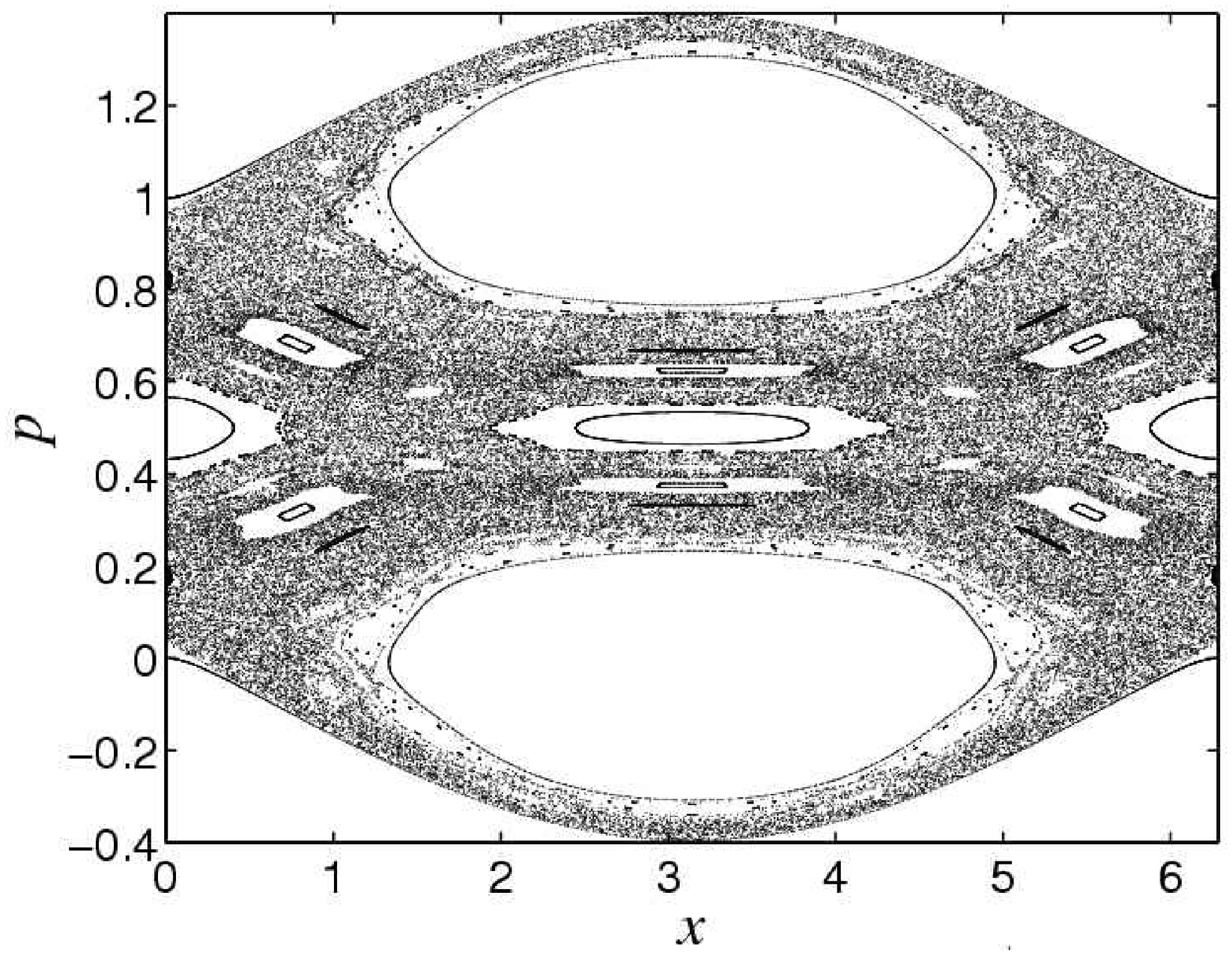,width=7.5cm,height=6.3cm}}
\put(3,3.9){\epsfig{file=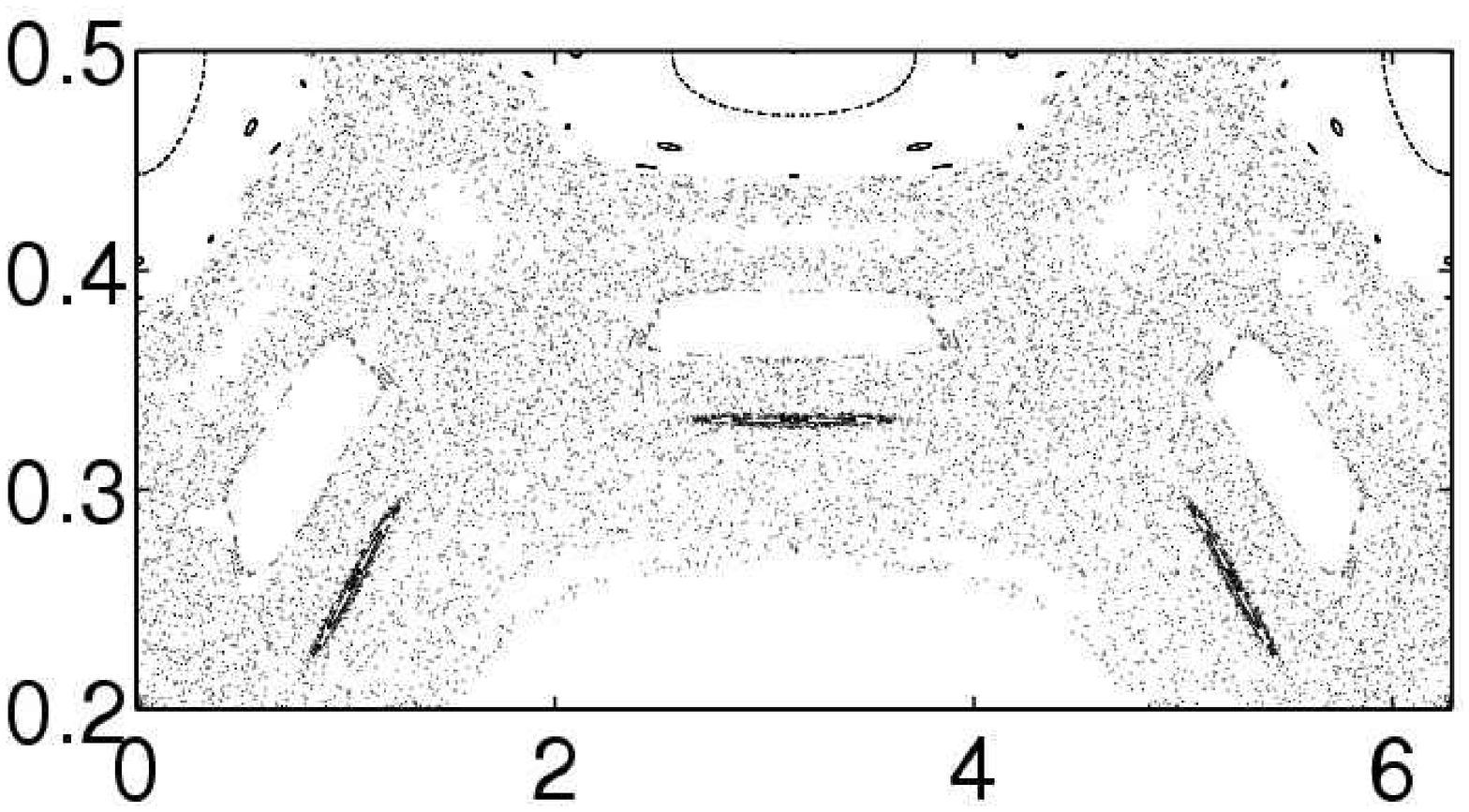,width=4.3cm,height=2.3cm}}
\end{picture}
\caption{Poincar\'e surface of section of Hamiltonian (\ref{eqn:fp}) with $\varepsilon=0.034$ (enlargement in the inset).}
\label{fig1}
\end{figure}

We consider the following forced pendulum model~:
\begin{equation}
\label{eqn:fp}
H(p,x,t)=\frac{1}{2}p^2+\varepsilon \left[ \cos x+\cos(x-t)\right].
\end{equation}
Figure~\ref{fig1} depicts a Poincar\'e section of Hamiltonian~(\ref{eqn:fp}) for $\varepsilon=0.034$. We notice that for $\varepsilon\geq 0.02759$ there are no longer any invariant rotational (KAM) torus~\cite{chanPR}.
First, this Hamiltonian with 1.5 degrees of freedom is mapped into an autonomous Hamiltonian with two degrees of freedom by considering that $t \mbox{ mod }2\pi$ is an additional angle variable. We denote $E$ its conjugate action.
The autonomous Hamiltonian is
\begin{equation}
\label{eqn:H2dof}
H=E+\frac{p^2}{2}+\varepsilon \left[\cos x+\cos(x-t)\right].
\end{equation}
The aim of the localised control is to modify locally Hamiltonian~(\ref{eqn:H2dof}) in order to reconstruct an invariant torus with frequency $\omega$. We assume that $\omega$ is sufficiently irrational in order to fulfill the hypotheses of the KAM theorem. First, the momentum $p$ is shifted by $\omega$ in order to define a localised control in the region $p\approx 0$ since the invariant torus is located near $p\approx\omega$ for Hamiltonian~(\ref{eqn:H2dof}) for $\varepsilon$ sufficiently small. The operators $\Gamma$ and ${\mathcal R}$ are defined from the integrable part of the Hamiltonian which is linear in the actions $(E,p)$~:
$$H_0(E,p)=E+\omega p,$$
and Hamiltonian~(\ref{eqn:H2dof}) is
$$
H=H_0+V,
$$ 
where 
\begin{equation}
\label{eqn:Vfp}
V(p,x,t)=\varepsilon \left[\cos x+\cos(x-t)\right]+\frac{p^2}{2}.
\end{equation}
The action of $\Gamma$, ${\mathcal R}$ and ${\mathcal N}$ on a function $U$ of $p$, $x$ and $t$ given by
$$
U(p,x,t)=\sum_{(k_1,k_2) \in {\mathbb Z}^2} U_{k_1,k_2}(p) {\mathrm e}^{i(k_1 x+k_2 t)},
$$
 are given by
\begin{eqnarray}
	&& \Gamma U =\sum_{(k_1,k_2)\not= (0,0)} \frac{U_{k_1,k_2}}{i(\omega k_1+k_2)} {\mathrm e}^{i(k_1 x+k_2 t)},\\
	&& {\mathcal R} U = U_{0,0}(p),\\
	&& {\mathcal N} U =\sum_{(k_1,k_2)\not= (0,0)}U_{k_1,k_2} {\mathrm e}^{i(k_1 x+k_2 t)}.
\end{eqnarray}

\subsubsection{Global control}

The actions of $\Gamma$, ${\mathcal R}$ and ${\mathcal N}$ on $V$ given by Eq.~(\ref{eqn:Vfp}) are
\begin{eqnarray*}
	&& \Gamma V =\varepsilon \Bigl[ \frac{\sin x}{\omega}+\frac{\sin (x-t)}{\omega-1}\Bigr],\\
	&& {\mathcal R} V=\frac{p^2}{2},\\
	&& {\mathcal N} V=\varepsilon \left[\cos x+\cos(x-t)\right].
\end{eqnarray*}
Since $\Gamma V$ depends only on $x$ and $t$, and since $V$ and ${\mathcal R}V$ are quadratic in $p$, it is straightforward to check that only the first two terms of the series~(\ref{eqn:ctf}) are non-zero. The global control term reduces to
$$
f(p,x,t)=-\frac{1}{2}\{\Gamma V\} ({\mathcal R}+1)V +\frac{1}{6} \{\Gamma V\}^2(2{\mathcal R}+1)V.
$$
Its explicit expression is given by
\begin{equation}
\label{eqn:ffp}
f(p,x,t)=\varepsilon p\left( \frac{\cos x}{\omega}+\frac{\cos(x-t)}{\omega-1}\right)+\frac{\varepsilon^2}{2}\left( \frac{\cos x}{\omega}+\frac{\cos(x-t)}{\omega-1}\right)^2.
\end{equation}
We notice that the control term is of order $\varepsilon$, i.e.\ of the same order as the perturbation. However, the control term $f$ acts only in a region where $\vert p\vert \lesssim \varepsilon$ since it is multiplied by a function $\Omega(\vert p\vert)$ such that $\Omega(\vert p\vert)=1$ when $\vert p\vert \leq \varepsilon$, and $\Omega(\vert p\vert)=0$ when $\vert p\vert \geq 2\varepsilon$. Consequently, the controlled Hamiltonian $H_0(E,p)+V(p,x,t)+f(p,x,t)\Omega(p) $ is locally integrable (since it is locally conjugate to $E+p^2/2$) provided that the canonical transformation is well defined (which is obtained when $\varepsilon$ is sufficiently small). A phase portrait of Hamiltonian~(\ref{eqn:fp}) with the control term~(\ref{eqn:ffp}) shows a very regular behaviour which persist for high values of $\varepsilon$. However we notice that for $\varepsilon$ greater than one, the control term is no longer small compared with the perturbation.

\subsubsection{Localised control}

In order to apply the localised control as in Sec.~\ref{sec:2b}, we notice that Hamiltonian~(\ref{eqn:H2dof}) is of the form~(\ref{eqn:Hcf}) with $v= \cos x+\cos(x-t)$, ${\bf w}={\bf 0}$ and $Q=p^2/2$. In this case the control term given by Eq.~(\ref{eqn:CTh}) is equal to
$$
f(x,t)=-\frac{1}{2}{\mathcal N} (\Gamma \partial_{x} v) ^2.
$$ 
Therefore the control term is equal to
\begin{equation}
	f(x,t)=-\frac{1}{2}\left( \frac{\cos x}{\omega}+\frac{\cos(x-t)}{\omega-1}\right)^2+\frac{1}{4}\left( \frac{1}{\omega^2}+\frac{1}{(\omega-1)^2}\right).
	\label{eqn:fpa}
\end{equation}
This control term has four Fourier modes with frequency vectors $(2,0)$, $(2,2)$, $(2,1)$ and $(0,1)$.
We consider the region in between the two primary resonances located around $p=0$ and $p=1$. The control term given by Eq.~(\ref{eqn:fpa}) can be simplified by considering the region of phase space around $p=1/2$. By keeping the main Fourier mode of this control term, i.e.\ the one with frequency vector $(2,1)$ which has the largest amplitude for $\omega$ close to $1/2$, the control term becomes~\cite{guido3,guido4}
\begin{equation}
f_a(x,t)=\frac{1}{2\omega(1-\omega)}\cos (2x-t).
\label{eqn:f2fpa}
\end{equation}
For the numerical computations we have chosen $\omega=(3-\sqrt{5})/2$ (golden-mean invariant torus) which is the last invariant torus to break-up for Hamiltonian~(\ref{eqn:fp}).

\begin{figure}
\epsfig{file=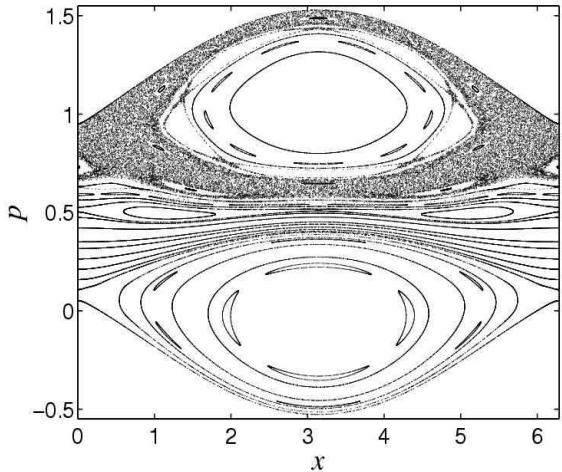,width=7.5cm,height=6.3cm}
\caption{Poincar\'e surface of section of Hamiltonian (\ref{eqn:fp}) with the global control term given by Eq.~(\ref{eqn:fpa}) with $\varepsilon=0.06$.}
\label{figCG}
\end{figure}

A Poincar\'e section of Hamiltonian~(\ref{eqn:fp}) with the approximate control term~(\ref{eqn:f2fpa}) for $\varepsilon=0.06$ shows that a lot of invariant tori are created with the addition of the control term precisely in the lower region of phase space where the localisation has been done (see Fig.~\ref{figCG}). Using the renormalization-group transformation~\cite{chanPR}, we have checked the existence of the golden-mean invariant torus for the Hamiltonian $H+\varepsilon^2 f$ is given by Eq.~(\ref{eqn:fpa}) with $\varepsilon\leq 0.06965$. By using the approximate and simpler control term $f_a$ given by Eq.~(\ref{eqn:f2fpa}) the existence of the invariant torus is obtained for $\varepsilon\leq 0.04857$. However, we have checked using Laskar's frequency map analysis~\cite{laskar} that invariant tori and effective barriers to diffusion (broken tori) persist up to higher values of the parameter ($\varepsilon \approx 0.2$).\\
The next step is to localize $f$ given by Eq.~(\ref{eqn:fpa}) around a chosen invariant torus created by $f$~: We assume that the controlled Hamiltonian~$H+\varepsilon^2 f$ has an invariant torus with the frequency $\omega$. We locate this invariant torus using frequency map analysis. Then we construct an approximation of the invariant torus of the Hamiltonian $H+\varepsilon^2 f$ of the form $p=p_0(x,t)$. We consider the following localised control term~:
\begin{equation}
\label{eqn:f2fpaL}
f^{(L)}(p,x,t)=f(x,t)\Omega(\vert p-p_0(x,t)\vert),
\end{equation}
where $\Omega$ is a smooth function with finite support around zero. More precisely, we have chosen $\Omega(x)=1$ for $x \leq \alpha$, $\Omega(x)=0$ for $x \geq \beta$ and a third order polynomial for $x \in ]\alpha,\beta[$ for which $\Omega$ is a $C^1$-function, i.e.\ $\Omega(x)=1-(x-\alpha)^2(3\beta-\alpha-2x)/(\beta-\alpha)^3$. The function $p_0$ and the parameters $\alpha$, $\beta$ are determined numerically ($\alpha=5\times 10^{-3}$ and $\beta=1.5\alpha$). The support in momentum $p$ of the localised control is of order $10^{-2}$ compared with the support of the global control which is of order 1.

\begin{figure}
\unitlength 1cm
\begin{picture}(15,6.3)(0,0)
\put(0,0){\epsfig{file=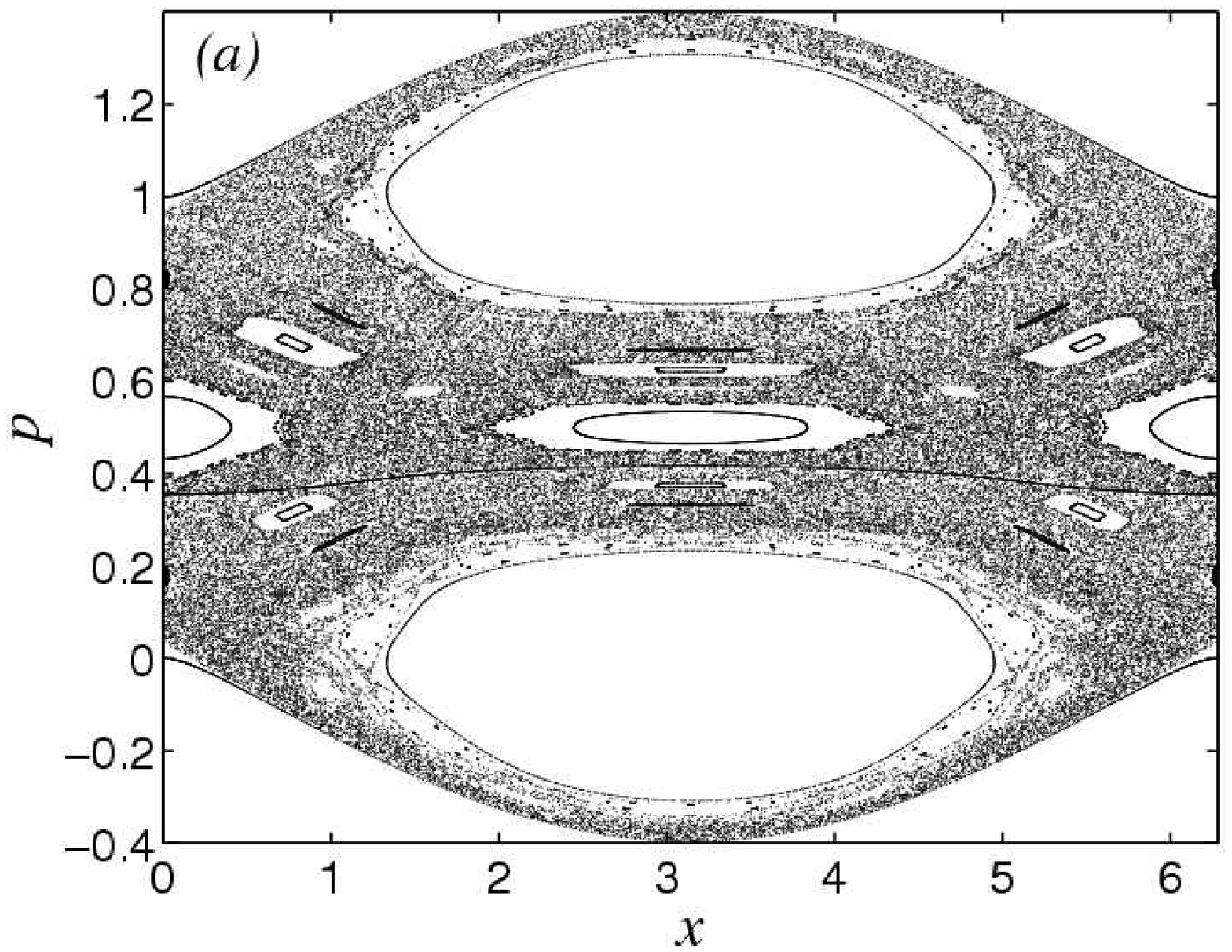,width=7.5cm,height=6.3cm}}
\put(3,3.9){\epsfig{file=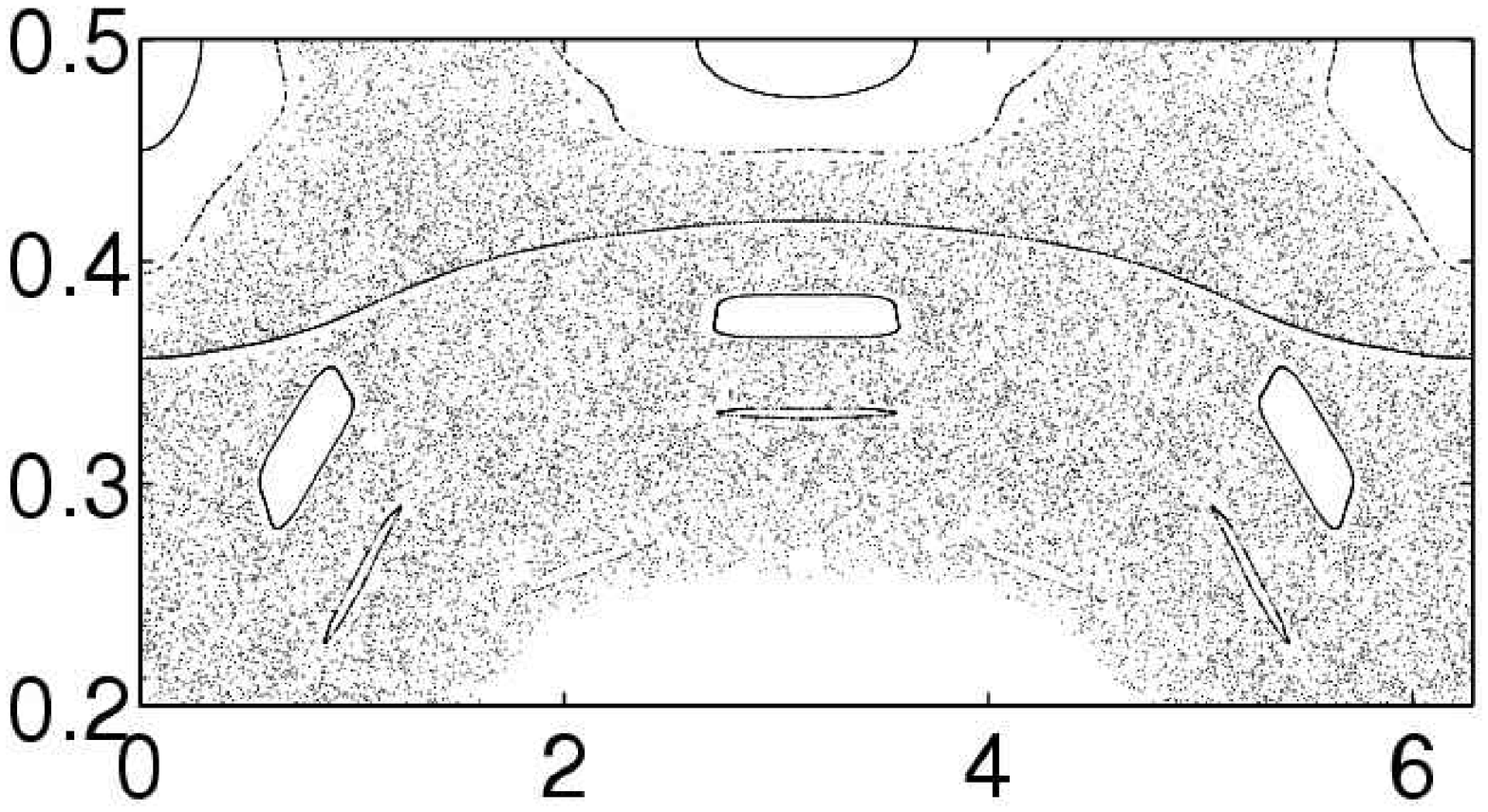,width=4.3cm,height=2.3cm}}
\put(8,0){\epsfig{file=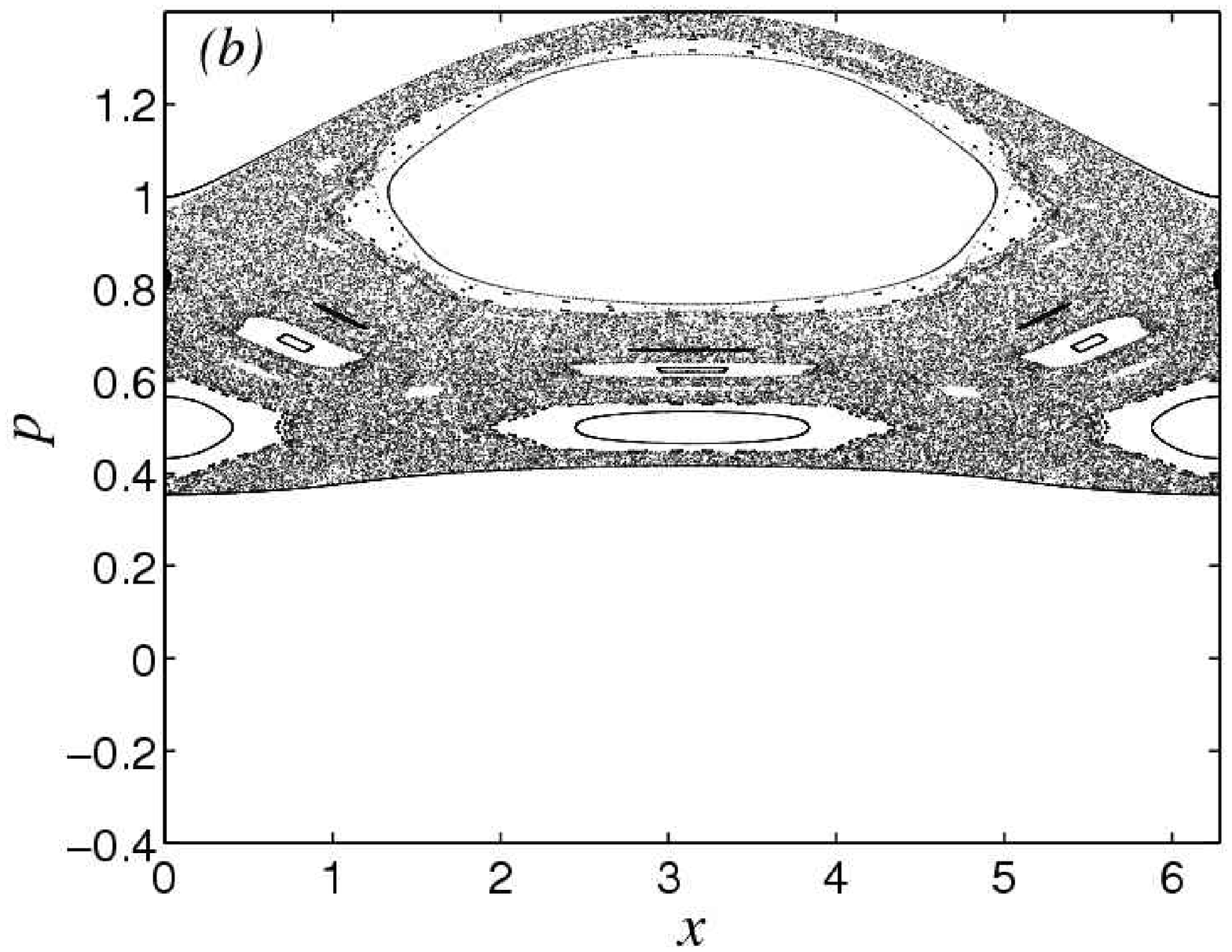,width=7.5cm,height=6.3cm}}
\end{picture}
\caption{$(a)$ Poincar\'e surface of section of Hamiltonian (\ref{eqn:fp}) with the approximate control term~(\ref{eqn:f2fpaL}) with $\varepsilon=0.034$ (enlargement in the inset). $(b)$ Same as $(a)$ with initial conditions above the invariant torus.}
\label{fig2}
\end{figure}

Figure~\ref{fig2} shows that the phase space of the controlled Hamiltonian is very similar to the one of the uncontrolled Hamiltonian. We notice that there is in addition an isolated invariant torus. Using frequency map analysis~\cite{laskar}, we check that this invariant torus corresponds to the one where the control term has been localised, i.e.\ its frequency is equal to $(3-\sqrt{5})/2$.

We notice that the perturbation has a norm (defined as the maximum of its amplitude) of $6.8\times 10^{-2}$ whereas the control term has a norm of $2.7\times 10^{-3}$ for $\varepsilon=0.034$. The control term is small (about 4\% ) compared to the perturbation. We notice that there is also the possibility of reducing the amplitude of the control (by a factor larger than 2) and still get an invariant torus of the desired frequency for a perturbation parameter $\varepsilon$ significantly greater than the critical value in the absence of control.

\subsection{Delta-kicked rotor -- standard map}

\begin{figure}
\epsfig{file=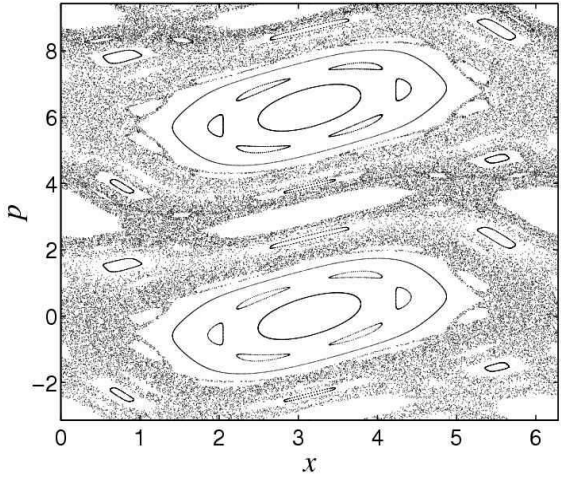,width=7.5cm,height=6.3cm}
\caption{Phase portrait of the standard map for $K=1.2$.}
\label{figSMc0}
\end{figure}

We consider the standard map~:
\begin{eqnarray}
	&& p_{n+1}=p_n+K\sin x_n,\\
	&& x_{n+1}=x_n+p_{n+1} \mbox{  mod } 2\pi.
\end{eqnarray}
This map is obtained by a Poincar\'e section of the following Hamiltonian
\begin{equation}
\label{eqn:Hsm}
H(p,x,t)=\frac{p^2}{2}+\frac{K}{4\pi^2}\sum_{m=-\infty}^{+\infty} \cos(x-mt).
\end{equation}
Figure~\ref{figSMc0} depicts a phase portrait of the standard map for $K=1.2$. We notice that there are no KAM tori (dividing phase space) at this value of $K$ (the critical value of the parameter for which all the KAM tori are broken is $K_c\approx 0.9716$). Similarly to the forced pendulum, we consider the invariant torus with frequency $\omega$. By translating the momentum, we map Hamiltonian~(\ref{eqn:Hsm}) into
\begin{equation}
\label{eqn:Hsml}
H=E+\omega p+\varepsilon \sum_{m=-\infty}^{+\infty}\cos (x-m t)+\frac{p^2}{2},
\end{equation}
where $\varepsilon=K/(4\pi^2)$. 

\subsubsection{Global control}
Using the same computations as for the forced pendulum, the controlled Hamiltonian obtained by the procedure described in Sec.~\ref{sec:2b} is 
\begin{eqnarray*}
H(p,x,t)=&&\frac{p^2}{2}+\varepsilon \sum_{m=-\infty}^{+\infty} \cos(x-mt)\\
&& +\varepsilon (p-\omega) \sum_{m=-\infty}^{+\infty} \frac{\cos (x-mt)}{\omega-m}+\frac{\varepsilon ^2}{2}\left( \sum_{m=-\infty}^{+\infty}\frac{\cos (x-mt)}{\omega-m}\right)^2,
\end{eqnarray*}
which is integrable and canonically conjugate to $H_0=p^2/2$. We notice that although the control term is of the same order as the perturbation, it is more regular than the perturbation since its Fourier coefficients decrease like $m^{-1}$. 

\subsubsection{Localised control}
Hamiltonian~(\ref{eqn:Hsml}) is of the form~(\ref{eqn:Hcf}) with $v=\sum_m \cos (x-mt)$, ${\bf w}={\bf 0}$ and $Q=p^2/2$.

The control term given by Eq.~(\ref{eqn:CTh}) becomes 
\begin{equation}
\label{eqn:fsma}
f(x,t)=-\frac{1}{2}\left(\sum_{m=-\infty}^{+\infty} \frac{\cos(x-mt)}{\omega-m}\right)^2+\frac{1}{4}
\sum_{m=-\infty}^{+\infty}\frac{1}{(\omega-m)^2}.
\end{equation}
Again we notice that this control term is more regular than the perturbation since its Fourier coefficients decrease like $m^{-1}$. In particular, it is bounded in space and time, piecewise continuous in time. For instance, for $\omega=1/2$, the control term (\ref{eqn:fsma}) is equal to
\begin{equation}
\label{eqn:SMT}
f(x,t)=\frac{\pi^2}{4}\cos(2x-t)\chi(t\notin 2\pi {\mathbb{Z}}) +\frac{\pi^2}{4}\chi(t\in 2\pi {\mathbb{Z}}).
\end{equation}

\begin{figure}
\unitlength 1cm
\begin{picture}(17,7)
\put(0,0){\epsfig{file=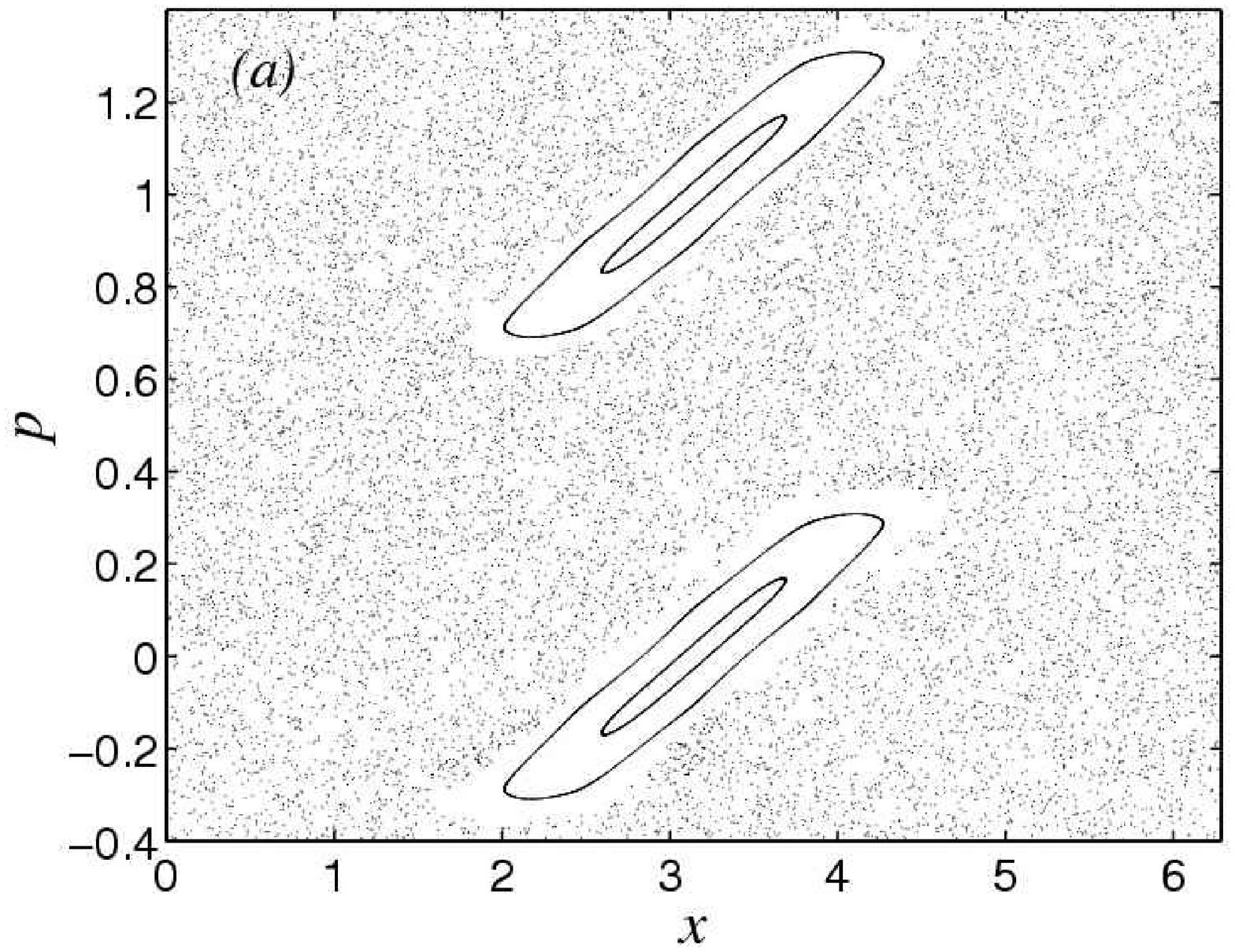,width=7.5cm,height=6.3cm}}
\put(8.5,0){\epsfig{file=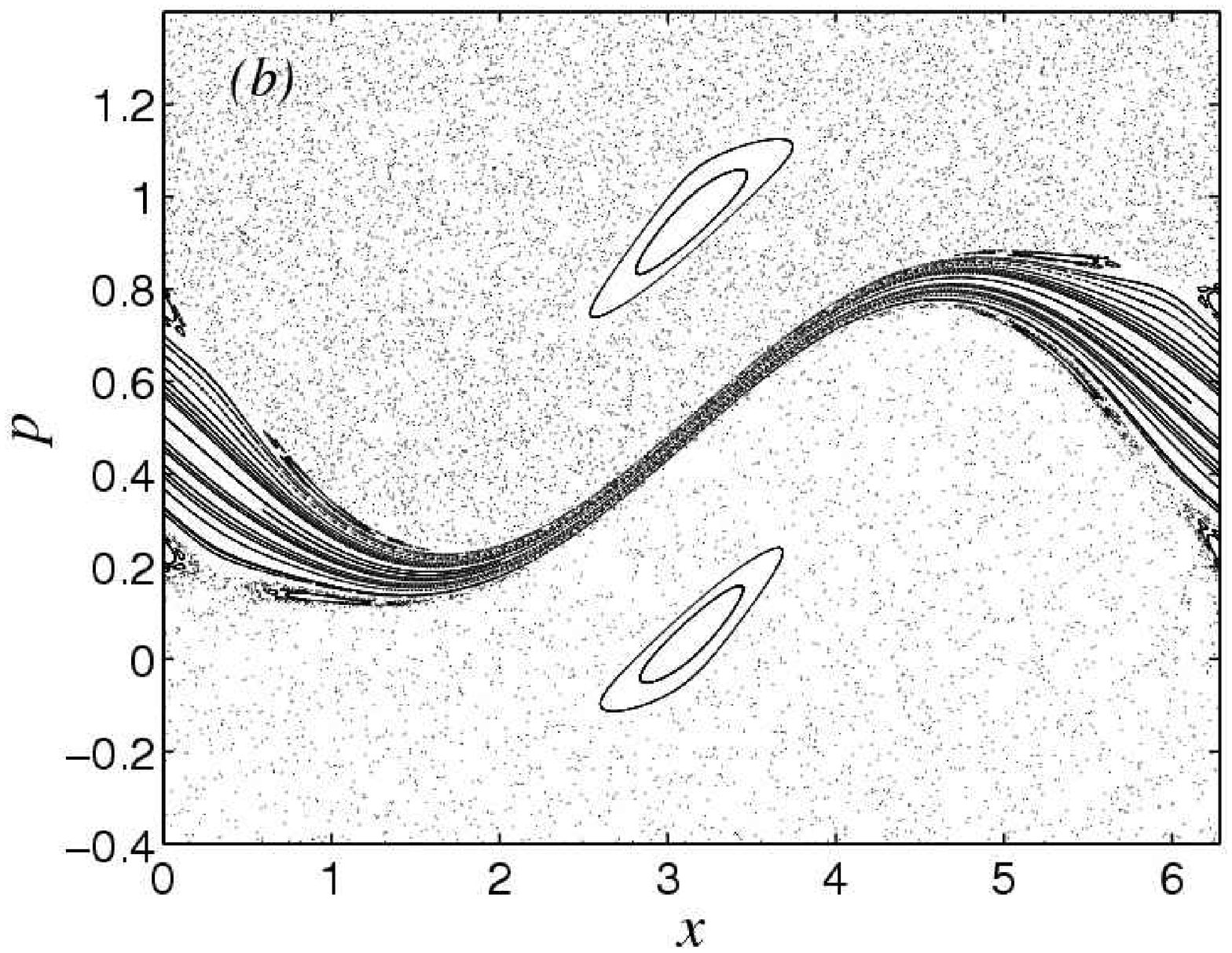,width=7.5cm,height=6.3cm}}
\end{picture}
\caption{Phase portrait of $(a)$ the standard map for $K=5$ and $(b)$ the controlled standard map with the control term~(\ref{eqn:SMT}).}
\label{figSMcT}
\end{figure}

The phase portrait of Hamiltonian~(\ref{eqn:Hsml}) with the control term~(\ref{eqn:SMT}) for $K=5$ is depicted on Fig.~\ref{figSMcT}. We notice that in this case, the controlled kicked rotor is now a kicked pendulum~: Instead of the rotor $H_0=p^2/2$, the integrable part becomes a pendulum
$$
H_0=\frac{p^2}{2}+\frac{\varepsilon^2\pi^2}{4}\cos(2x-t),
$$
and the perturbation is a periodic $\delta$-kick. 
We notice that the controlled Hamiltonian has invariant tori in the region near $p=1/2$ (where the control has been localised). These invariant tori persist up to high values of the parameter $K=\varepsilon/(4\pi^2)$ larger than 10 which has to be compared with $K_c\approx 0.97$ in the absence of control. Note that the control term we use is bounded (conversely to the perturbation) and its amplitude is small compared with the amplitude of the Fourier coefficients of the perturbation (with a factor smaller than 10\% depending on $K$). 

In order to recover a map, we need to locate the control at each $\omega=(2m+1)/2$ for $m\in {\mathbb Z}$. 
For a given $m\in{\mathbb Z}$, the approximate control term is
$$
f^{(a)}_m(x,t)=\frac{\pi^2}{4}\cos(2x-(2m+1)t).
$$
Since each of these control term act on different regions of phase space, we sum all these control term to obtain a more global stabilization and recover a map.
The control term becomes
$$
f^{(a)}(x,t)=\frac{\pi^2}{4}\sum_{m=-\infty}^{+\infty} \cos(2x-(2m+1)t).
$$
Next, we perform an inverse Fourier transform. The Hamiltonian becomes
$$
f^{(a)}(x,t)=\frac{\pi^3}{4}\cos 2 x \sum_{m=-\infty}^{+\infty}\left[ \delta (t-2\pi m) -\delta (t-(2m+1)\pi)\right].
$$
The controlled standard map is thus obtained by performing additional kicks: Each $t=2\pi m$, in addition to the kicks of strength $K\cos x$, one has to perform kicks of strength $(K^2\cos 2 x)/16$, and each $t=(2m+1)\pi$, one has to perform kicks of strength $-(K^2\cos 2 x)/16$. This leads to the following form for the controlled standard map
\begin{eqnarray*}
	&& \tilde{p}_{2m+1}=\tilde{p}_{2m}+K\sin \tilde{x}_{2m}+\frac{K^2}{16}\sin 2\tilde{x}_{2m},\\
	&& \tilde{p}_{2m+2}=\tilde{p}_{2m+1}-\frac{K^2}{16}\sin 2\tilde{x}_{2m+1},\\
	&& \tilde{x}_{2m+1}=\tilde{x}_{2m}+\frac{1}{2}\tilde{p}_{2m+1},\\
	&& \tilde{x}_{2m+2}=\tilde{x}_{2m+1}+\frac{1}{2}\tilde{p}_{2m+2}.
\end{eqnarray*}
It has a more compact form by using $x_n=\tilde{x}_{2n}$ and $p_n=\tilde{p}_{2n}$~: 
\begin{eqnarray}
\label{eqn:smct1}
	&& p_{n+1}=p_n+K\sin x_n+\frac{K^2}{16}\bigl[ \sin 2x_n \nonumber \\
	&& \qquad \qquad -\sin (2x_n+p_n+K\sin x_n+(K^2\sin 2x_n)/16)\bigr],\\
	&& x_{n+1}=x_n+p_{n+1}+ \frac{K^2}{32}\sin (2x_n+p_n+K\sin x_n+(K^2\sin 2x_n)/16).
	\label{eqn:smct1b}
\end{eqnarray}
If we neglect the kicks every $t=(2m+1)\pi$, we obtain the following map
\begin{eqnarray}
\label{eqn:smct2}
	&& p_{n+1}=p_n+K\sin x_n +\frac{K^2}{16}\sin 2 x_n,\\
	&& x_{n+1}=x_n+p_{n+1}.\label{eqn:smct2b}
\end{eqnarray}
Conversely, if we neglect the kicks every $t=2m\pi$, we get
\begin{eqnarray}
\label{eqn:smct3}
	&& p_{n+1}=p_n+K\sin x_n -\frac{K^2}{16}\sin(2x_n+p_n+K\sin x_n),\\
	&& x_{n+1}=x_n+p_{n+1}+\frac{K^2}{32}\sin(2x_n+p_n+K\sin x_n).
	\label{eqn:smct3b}
\end{eqnarray}

\begin{figure}
\epsfig{file=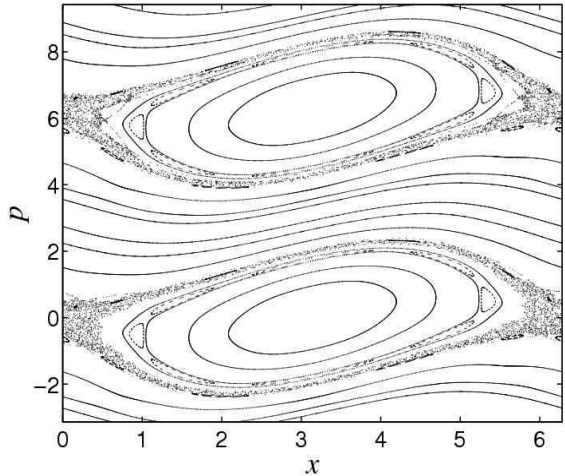,width=7.5cm,height=6.3cm}
\caption{Phase portrait of the controlled standard map~(\ref{eqn:smct1})-(\ref{eqn:smct1b}) for $K=1.2$.}
\label{figSMc1}
\end{figure}

\begin{figure}
\epsfig{file=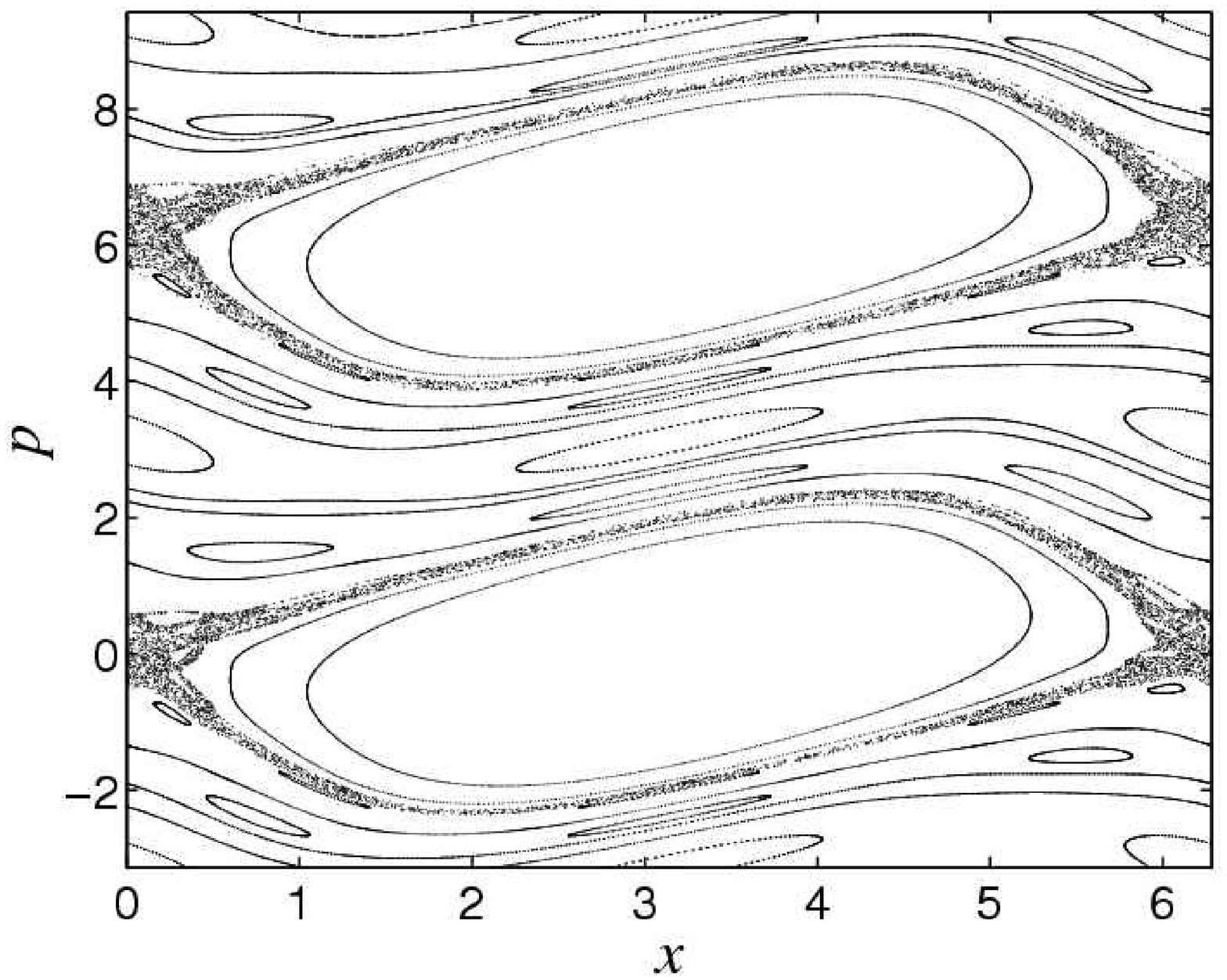,width=7.5cm,height=6.3cm}
\caption{Phase portrait of the controlled standard map~(\ref{eqn:smct2})-(\ref{eqn:smct2b}) for $K=1.2$.}
\label{figSMc2}
\end{figure}

\begin{figure}
\epsfig{file=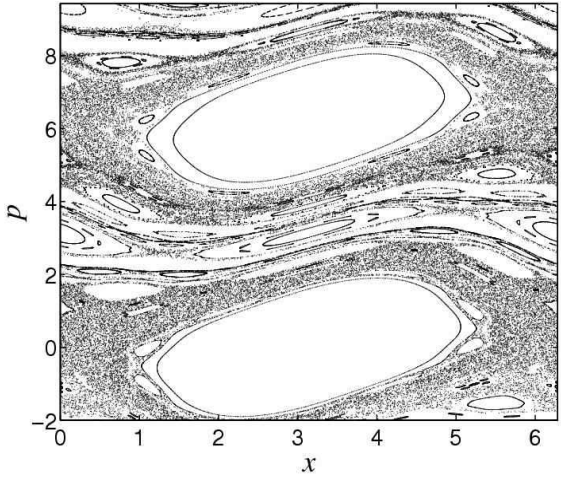,width=7.5cm,height=6.3cm}
\caption{Phase portrait of the controlled standard map~(\ref{eqn:smct3})-(\ref{eqn:smct3b}) for $K=1.2$.}
\label{figSMc3}
\end{figure}

A phase portrait of these maps are depicted on Figs.~\ref{figSMc1}, \ref{figSMc2} and \ref{figSMc3}. The most efficient control is obtained for the map~(\ref{eqn:smct1})-(\ref{eqn:smct1b}) by considering the two additional kicks of order $K^2$. The control term which only add the negative kicks does not lead to an efficient control although it appears slightly more regular than the uncontrolled case. Using frequency map analysis~\cite{laskar}, we have computed the critical thresholds for the break-up of the last KAM tori~: There are invariant tori for the map~(\ref{eqn:smct1})-(\ref{eqn:smct1b}) up to $K_c\approx 2.23$ which is more than twice the uncontrolled case ($K_c\approx 0.97$). The map~(\ref{eqn:smct2})-(\ref{eqn:smct2b}) which is simpler than the map~(\ref{eqn:smct1})-(\ref{eqn:smct1b}) has invariant tori up to $K_c\approx 1.57$.

In this section, we have used the control for Hamiltonian flows in order to derive control terms for area-preserving maps. We note that a control method has been developed directly for area-preserving maps in Refs.~\cite{cvecp}. 

\subsection{Non-twist Hamiltonian}

\begin{figure}
\epsfig{file=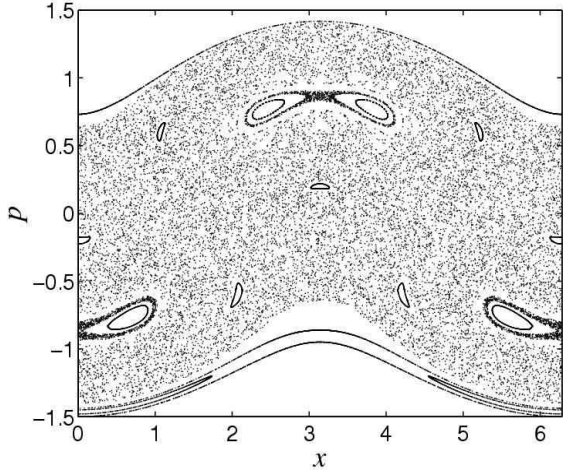,width=7.5cm,height=6.3cm}
\caption{Poincar\'e section for Hamiltonian~(\ref{eqn:Hcub}) with $\varepsilon=0.2$ and $\omega=(\sqrt{5}-1)/2$.}
\label{fig:cub}
\end{figure}

\begin{figure}
\epsfig{file=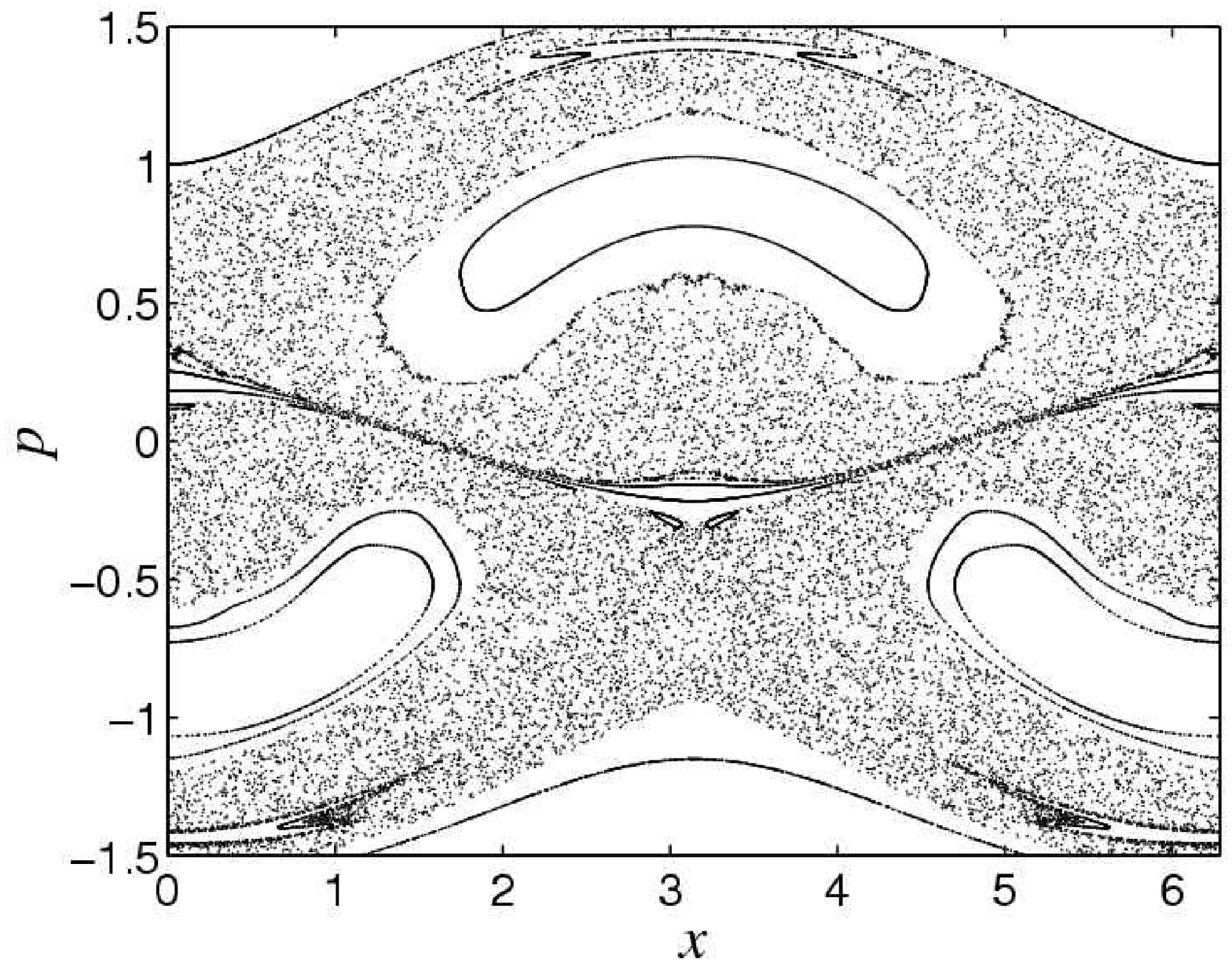,width=7.5cm,height=6.3cm}
\caption{Poincar\'e section for Hamiltonian~(\ref{eqn:Hcub}) with the control term~(\ref{eqn:HcubC}) with $\varepsilon=0.2$ and $\omega=(\sqrt{5}-1)/2$.}
\label{fig:cubC}
\end{figure}

We consider the following Hamiltonian
\begin{equation}
\label{eqn:Hcub}
H=E+\omega p +\frac{p^3}{3}+\varepsilon \left( \cos x +\cos (x-t)\right),
\end{equation}
where $\omega=(\sqrt{5}-1)/2$. A Poincar\'e section of this Hamiltonian with $\varepsilon=0.2$ is depicted on Fig.~\ref{fig:cub}.
The invariant torus with frequency $\omega$ is located at $p=0$ for $\varepsilon=0$. We notice that this invariant torus is shearless since the second derivative of $H$ with respect to $p$ is zero on this torus. Hamiltonian~(\ref{eqn:Hcub}) is of the form given by Eq.~(\ref{eqn:Hcf}) with $v=\cos x +\cos (x-t)$, ${\bf w}={\bf 0}$ and $Q=p^3/3$. The control term is given by Eq.~(\ref{eqn:CTh}):
\begin{equation}
\label{eqn:HcubC}
\varepsilon^2 f(x,t)=\frac{\varepsilon^3}{3}\left(\frac{\cos x}{\omega}+\frac{\cos (x-t)}{\omega-1}\right)^3.
\end{equation}
The noticeable feature is that the modification of the Hamiltonian is of order $\varepsilon^3$ compared with the forced pendulum or the standard map where the control term is of order $\varepsilon^2$. A Poincar\'e section of the controlled Hamiltonian~(\ref{eqn:Hcub}) with the control term (\ref{eqn:HcubC}) is depicted on Fig.~\ref{fig:cubC}. We notice that there are invariant tori in the region near $p=0$ that have been created with the addition of the small control term of order $\varepsilon^3$. For instance, for $\varepsilon=0.2$ and $\omega=(\sqrt{5}-1)/2$, the control term has a norm which is about 10\% of the perturbation. In order to obtain a localised control, one has to apply the control term only in the region where the KAM invariant tori have been recreated, i.e., in the region $p\approx 0$.

\ack
We acknowledge useful discussions with Ph Ghendrih, M Pettini and V Zagrebnov.
This work is supported by Euratom/CEA 
(contract $V3382.001$).

\appendix

\section{The transformations $T_{{\bf m},{\bf b}}$ are canonical}
\label{sec:appa}

First we notice that $T_{{\bf m},{\bf b}}$ can be written as
$$
T_{{\bf m},{\bf b}}={\mathrm e}^{-{\bf m}\partial }{\mathcal D}_{1+\kM'} {\mathcal T}_{\bf b},
$$
where ${\mathcal D}$ is a dilatation and ${\mathcal T}$ a translation of the actions acting on a function $V({\bf A},{\bm\theta})$ as
\begin{eqnarray*}
	&& ({\mathcal D}_{1+\kM'} V)({\bf A},{\bm\theta})=V(({1+\kM'}) {\bf A},{\bm\theta}),\\
	&& ({\mathcal T}_{\bf b} V)({\bf A},{\bm\theta})=V({\bf A}+{\bf b},{\bm\theta}).
\end{eqnarray*}

A translation of the actions by a function ${\bf b}$ of ${\bm\theta}$ (i.e.\ independent of ${\bf A}$) is obviously a canonical transformation if ${\bf b}$ derives from a scalar function, i.e.\ ${\bf b}=\partial \beta$. Thus it is sufficient to prove that $T_{{\bf m},{\bf 0}}$ is also a canonical transformation. We notice that $T_{{\bf m},{\bf 0}}$ is an automorphism since it is the product of two automorphisms~: an exponential of a derivation and a dilatation. In what follows we prove that
\begin{equation}
\label{eqn:eqap}
{\mathrm e}^{-{\bf m}\partial }{\mathcal D}_{1+\kM'}={\mathrm e}^{\{ {\bf m}{\bf A}\}},
\end{equation}
which is equivalent to say that $T_{{\bf m},{\bf 0}}$ is a Lie transform generated by the scalar function $ {\bf m}{\bf A}$. In other terms, Eq.~(\ref{eqn:eqap}) can also be written as [cf. Eq.~(\ref{eqn:wf})]
$$
{\mathcal D}_{{\mathrm e}^{\partial\hat{\bf m}}\cdot 1}={\mathrm e}^{{\bf m}\partial} {\mathrm e}^{\{{\bf m}{\bf A}\}}.
$$
Since these operators are automorphisms, it is sufficient to check that Eq.~(\ref{eqn:eqap}) is satisfied on the basis $({\bf A},{\bm\theta})$. First, we expand the operator $ {\mathrm e}^{\{ {\bf m}{\bf A}\}}$~:
$$
{\mathrm e}^{\{ {\bf m}{\bf A}\}}={\mathrm e}^{-{\bf m}\partial +\overline{{\bf m}'{\bf A}}\partial_{\bar{\bf A}}},
$$
and we use the Trotter-Kato formula~\cite{TK} to express the exponential of the sum of two operators~:
$$
{\mathrm e}^{{\mathcal A}+{\mathcal B}}=\lim_{n\to \infty }\left( {\mathrm e}^{{\mathcal A}/n}{\mathrm e}^{{\mathcal B}/n}\right)^n.
$$
Any function of ${\bm\theta}$ is invariant under the action of an operator ${\mathrm e}^{\overline{{\bf m}'{\bf A}}\partial_{\bar{\bf A}}/n }$. Therefore it is straightforward to check that
$$
\left( {\mathrm e}^{-{\bf m}\partial  /n}{\mathrm e}^{ \overline{{\bf m}'{\bf A}}\partial_{\bar{\bf A}} /n}\right)^n {\bm \theta} = {\mathrm e}^{-{\bf m}\partial }{\bm\theta},
$$
for all $n\in {\mathbb N}$. It follows that Eq.(\ref{eqn:eqap}) is satisfied on ${\bm\theta}$. 

For ${\bf A}$, we use the identity
\begin{equation}
\label{eqn:ema}
{\mathrm e}^{\overline{M {\bf A}}\partial_{\bar{\bf A}}} N {\bf A}=N{\mathrm e}^{M} {\bf A},
\end{equation}
where $M$ and $N$ are functions of $\bm\theta$. This identity follows from $(\overline{M{\bf A}} \partial_{\bar{\bf A}})N {\bf A}= N M {\bf A}$. In particular, we have
$$
{\mathrm e}^{\overline{M {\bf A}}\partial_{\bar{\bf A}}}={\mathcal D}_{{\mathrm e}^M}.
$$
Using Eq.~(\ref{eqn:ema}) we prove that
$$
\left( {\mathrm e}^{-{\bf m}\partial   /n}{\mathrm e}^{ \overline{{\bf m}'{\bf A}}\partial_{\bar{\bf A}} /n}\right) {\bf A}={\mathrm e}^{-{\bf m}\partial /n}{\mathrm e}^{ {\bf m}'/n} {\bf A}, 
$$
and it follows recursively that
$$
\left( {\mathrm e}^{-{\bf m}\partial   /n}{\mathrm e}^{ \overline{{\bf m}'{\bf A}}\partial_{\bar{\bf A}} /n}\right)^n {\bf A}=\prod_{k=1}^n {\mathrm e}^{-k{\bf m}\partial /n}\left( {\mathrm e}^{{\bf m}'/n}\right) {\bf A},
$$
where the product is taken from right to left, i.e.\ $\prod_{k=1}^n a_k=a_n a_{n-1}\cdots a_1$.
Concerning the operator ${\mathrm e}^{-{\bf m}\partial}{\mathcal D}_{1+\kM'}$, the Trotter-Kato formula leads to the following expansion
$$
{\mathrm e}^{-{\bf m}\partial }{\mathcal D}_{1+\kM'}=
\lim_{n\to\infty} {\mathrm e}^{-{\bf m}\partial } \left( {\mathrm e}^{{\bf m}'/n}{\mathrm e}^{{\bf m}\partial /n}\right)^n {\bf A},
$$
since $ 1+\kM'={\mathrm e}^{\partial  \hat{{\bf m}}}\cdot 1$ and $\partial  \hat{{\bf m}}={\bf m}' +{\bf m}\partial $ [see Eq.~(\ref{eqn:dmu})].
Using the same type of computations as for $T_{{\bf m},{\bf 0}}$, we have
$$
\left( {\mathrm e}^{{\bf m}'/n}{\mathrm e}^{{\bf m}\partial /n}\right)^n=
\prod_{k=n-1}^0 {\mathrm e}^{k{\bf m}\partial /n}\left({\mathrm e}^{{\bf m}'/n}\right).
$$
If we multiply the above expression by ${\mathrm e}^{-{\bf m}\partial }$ and change the index of the product ($k'=n-k$), it leads to 
$$
{\mathrm e}^{\{ {\bf m}{\bf A}\}} {\bf A}= {\mathrm e}^{-{\bf m}\partial}{\mathcal D}_{1+\kM'} {\bf A},
$$
and hence Eq.~(\ref{eqn:eqap}) is proved.\\

{\em Remark~:} We notice that Eq.~(\ref{eqn:eqap}) implies that 
$$
\left( {\mathcal D}_{{\mathrm e}^{\partial \hat{\bf m}}\cdot 1}\right)^{-1}={\mathrm e}^{{\bf m}\partial }
{\mathcal D}_{{\mathrm e}^{-\partial \hat{\bf m}}\cdot 1}{\mathrm e}^{-{\bf m}\partial },
$$
which leads to the expression of the inverse of $1+\kM'$~:
$$
\left( {\mathrm e}^{\partial \hat{\bf m}}\cdot 1\right)^{-1}={\mathrm e}^{{\bf m}\partial } {\mathrm e}^{-\partial \hat{\bf m}}\cdot 1.
$$

\end{document}